\documentclass[epj,final]{svjour}
\usepackage{graphics}
\usepackage{graphicx,bm}
\usepackage{rotating}
\usepackage{multirow}
\usepackage{array}
\usepackage{amsmath}
\usepackage{amssymb}

\begin{document}
\title{Efimov physics beyond universality}

\author{Richard Schmidt \and Steffen Patrick Rath \and Wilhelm Zwerger 
}                     
%
%
\institute{Physik Department, Technische Universit\"at M\"unchen,
  85747 Garching, Germany}
\date{Received: date / Revised version: date}
%
\abstract{
We provide an exact solution of the Efimov spectrum in ultracold 
gases within the standard two-channel model for Feshbach resonances.
 It is shown that the finite range in the Feshbach coupling 
 makes the introduction of an adjustable three-body parameter obsolete.
The solution explains the empirical relation between the scattering length 
$a_{-}$ where the first Efimov state appears at the atom threshold and the van der 
Waals length $l_\text{vdw}$ for open-channel dominated resonances.
There is a continuous crossover to the closed-channel dominated limit,
where the scale in the energy level diagram as a function of the inverse 
scattering length $1/a$ is set by the intrinsic length $r^{\star}$ associated with the Feshbach coupling.
Our results provide a number of predictions for the deviations from universal scaling relations between
energies and scattering lengths that can be tested in future experiments. 
}
%


\maketitle
\section{Introduction}

Most of the basic features that distinguish quantum from classical physics 
show up already at the single particle level. Genuine two-particle effects
like the Hong-Ou-Man-del two-photon interference are typically a consequence 
of particle statistics, not of interactions \cite{Hong1987}. Surprisingly, novel quantum effects in which
statistics and interactions are combined appear at the level of three particles. As
shown by Efimov in 1970~\cite{Efimov1970}, three particles which interact via a resonant short-range 
attractive interaction exhibit an infinite sequence of three-body bound states
or trimers.  Remarkably,  the trimers exist even in a regime where the two-body interaction 
does not have a bound state.  Efimov trimers thus behave like 
Borromean rings: three of them are bound together but cutting one of the bonds 
makes the whole system fly apart. 
While theoretically predicted in a nuclear matter context, Efimov states have finally been observed with ultracold atoms \cite{Kraemer2006}. 
The assumption of short range interactions is perfectly valid in this case and, moreover, the associated scattering 
lengths can be tuned by an external magnetic field, exploiting a Feshbach resonance~\cite{Chin2010}.  An important feature of the Efimov trimers is that the binding energies exhibit universal scaling behavior. In the limit where the two-body interaction is just at the threshold to form a bound state, the ratio $E^{(n)}/E^{(n+1)}$ of consecutive binding energies approaches the universal value $e^{2\pi/s_0}\simeq 515.028$ for $n\gg 1$ with the Efimov number  $s_0\approx1.00624$.  One key experimental signature of Efimov physics is the resonant enhancement of the three-body recombination rate when the $n$th Efimov state meets the atom threshold at a scattering length $a^{(n)}_{-}$. Also here the universal scaling law prevails: the ratio of consecutive values of $a^{(n)}_{-}$ approaches $a_{-}^{(n+1)}/a_{-}^{(n)}\!\to\! e^{\pi/s_0}\simeq 22.6942$. The origin of this universality can be understood from an effective field theory approach to the three-body problem
~\cite{Bedaque1999a,Braaten2006}. There remains, however, a non-universal aspect in the theory: Although the relative position of the trimer states is universal, this does not fix their absolute position in the $(a,E)$ plane which is determined by the so-called three-body parameter (3BP). It is presumed that the 3BP is highly sensitive to microscopic details of the underlying two-body potential as well as genuine three-body forces~\cite{DIncao2009}.\\
 In practice, it is often only the lowest Efimov state at $a_{-}\! =a_{-}^{(0)}$ that can be observed because of large atom losses as the scattering length increases. As more experimental data have been accumulated in recent years \cite{Ottenstein2008,Huckans2009,Pollack2009,Zaccanti2009,Gross2009,Gross2010,Berninger2011,Wild2011}, a puzzling observation came to light: In most experiments, the measured values for $a_{-}$ clustered around $a_{-}\approx -9.45\, l_{\text{vdw}}$, no matter which alkali atoms were used. Since $a_{-}$ determines the overall scale of the whole Efimov spectrum,  this observation suggests a three-body parameter which is independent of the microscopic details. But where does this apparent `universality of the three-body parameter' come from? A possible answer to this question is based on the observation that, typically, Efimov trimers which are accessible with ultracold atoms appear in a situation where the scattering length is tuned via an open-channel dominated Feshbach resonance. Such resonances are well described by a single-channel picture. Irrespective of the short distance behavior, the associated  \textit{two}-body problem is then known to 
have $l_\text{vdw}$ as the only relevant length scale at energies much smaller than the depth of the potential well~\cite{Flambaum1999}. In the absence of genuine three-body forces, it is plausible that this result extends to the three-particle Efimov problem and thus,
that $l_\text{vdw}$ provides the characteristic scale for the 3BP. This has in fact been shown in recent, independent  work on this problem by 
Chin~\cite{Chin2011} and by Wang \textit{et al.}~\cite{Greene2012}, using single-channel potentials with a van der Waals tail. 

While being consistent with the observed correlation between  $a_{-}$ and the van der Waals length $l_\text{vdw}$ for a number of different alkalis,
such a single-channel description does not apply in general and thus suggests a universality of the ratio $a_{-}/l_\text{vdw}$ which is far too general 
even within the constraint that only two-body interactions play a role. In particular, it fails for closed-channel dominated Feshbach resonances. 
In fact, in this case, it is known that the 3BP is set by  the intrinsic length $r^{\star}$ which determines the strength of the Feshbach 
coupling~\cite{Petrov2004,Gogolin2008}. In the following, we present an exact solution for the Efimov spectrum within a standard 
two-channel model~\cite{Chin2010} which incorporates the finite range of the Feshbach 
coupling and properly recovers both limits of open-channel \textit{and} closed-channel dominated resonances.  It provides a complete 
description of the trimer spectrum in terms of only two, experimentally accessible, parameters: the van der Waals length $l_{\text{vdw}}$ and the intrinsic length $r^{\star}$.  Depending on the dimensionless resonance strength $s_{\rm res}=0.956\, l_\text{vdw}/r^{\star}$ \cite{Chin2010}, 
there is a continuous change in the relation between the trimer energy spectrum and the scattering length,
with the lowest Efimov state appearing at $a_-\approx -8.3\, l_\text{vdw}$ as  $s_{\rm res}\gg 1$ while $a_-\approx -10.3\, r^{\star}$ in 
the opposite limit $s_{\rm res} \ll 1$. 
This model provides a minimum description of the Efimov spectrum which is based on two-body physics only and has no adjustable parameter. It explains why the ratio $a_{-}/l_\text{vdw}$ is in the observed range for open-channel dominated resonances and 
predicts strong deviations from this in the intermediate regime $s_{\rm res}=\mathcal{O}(1)$. 
Our results are consistent with most of the experimental data, even though no details of the interatomic potentials at short  distances or three-body 
forces are included. As an important additional feature, we find that the 
experimentally accessible lowest Efimov states exhibit strong deviations from the universal ratios that characterize the 
scaling limit,  which have apparently been observed in recent experiments \cite{Wenz2009,Williams2009}. 

\section{Two-channel model}
We consider non-relativistic bosons described by the microscopic action (in units where $2m=\hbar=1$)
\begin{eqnarray}\label{eq:action}
S&=&\int_{\mathbf r,t}\Big\{ \psi^*(\mathbf r,t) [i \partial_t-\nabla^2] \psi(\mathbf r,t) \nonumber\\
&+&\, \phi^*(\mathbf r,t) P_\phi^\text{cl}\phi(\mathbf r,t)  \Big\}+\frac{g}{2} \int_{\mathbf r_1,\mathbf r_2,t} \chi(\mathbf r_2-\mathbf r_1) \times\nonumber\\
&&\Big[ \phi(\frac{\mathbf r_1+\mathbf r_2}{2},t)
 \psi^*(\mathbf r_1,t) \psi^*(\mathbf r_2,t) + c. c. \Big],
\end{eqnarray}
where $\psi$ denotes the atoms and $\phi$ the molecule in the closed channel. Here $P_\phi^\text{cl}= i \partial_t-\nabla^2/2+\nu$ with 
$\nu(B)=\mu(B-B_\text{res})$ the bare detuning from the resonance and $\mu$ is the difference in the magnetic moment between the molecule  and the open-channel atoms.  For a description of universal features of Efimov trimers like the asymptotic ratio  $a_-^{(n+1)}/a_-^{(n)}\!\to\! e^{\pi/s_0}$, the 
atom-molecule conversion amplitude $\sim g$ may be taken as pointlike in coordinate space~\cite{Bedaque1999a,Braaten2006}.
In reality, however, the coupling has a finite range $\sigma$ which is determined by the scale of the wave function overlap 
between the open- and closed-channel states. As has been pointed out by a number of 
authors~\cite{Szymanska2005,Massignan2008,Werner2009,Pricoupenko2010,JonaLasinio2010,Pricoupenko2011}, this can be accounted for 
by a form factor $\chi(r)$ in Eq. \eqref{eq:action}. The solution of the modified Skornyakov-Ter-Martirosian (STM) equation \eqref{STM} below, can then be used to fit the location of three-body resonances for different alkali atoms by adjusting the associated range parameter $\sigma$ \cite{Pricoupenko2010,JonaLasinio2010,Pricoupenko2011}. The precise form of the form factor $\chi(r)$ depends on details of the interatomic potentials. As will show below, however, its characteristic length 
$\sigma$ is given by the van der Waals or the mean scattering length $\bar a$. In physical terms, this reflects the fact that 
the classical turning point in the closed-channel states is of the order of $l_\text{vdw}$ because for typical magnetic field-tuned Feshbach resonances, 
it is only the bound states close to the continuum threshold that are experimentally accessible. Specifically, we choose an exponential form factor 
$\chi(r)\sim e^{-r/\sigma}/r$, which leads to $\chi(p)=1/(1+\sigma^2 p^2)$ in momentum space. In contrast to the more standard Gaussian 
cutoff~\cite{Szymanska2005,Massignan2008,Werner2009}, this choice is in fact optimal in the sense that the resulting effective range 
$r_e=3\,\bar a$ of two-body scattering near an open-channel dominated Feshbach resonance (see Eq. \eqref{effrange} below) agrees very well with the standard result
$r_e\approx 2.92\, \bar a$~\cite{Flambaum1999} for a single-channel potential with a $1/r^6$ tail.
It is important to note that the action \eqref{eq:action} can also be used to describe the situation where the interaction is 
dominated by a large background scattering length $a_\text{bg}$. Indeed, integrating out the closed-channel field $\phi$,
one obtains a contribution $\sim(\psi^*\psi)^2$ that properly describes background scattering of range $\sigma$ 
and scattering length $\sim g^2/\tilde\nu$,  provided the Feshbach coupling $g^2=32\pi/r^{\star}\gg 1/l_\text{vdw}$ is strong 
enough such that the momentum dependence of $P_\phi^\text{cl}$ can be neglected. In the opposite limit of closed-channel dominated resonances,
however, using a proper two-channel model cannot be avoided~\cite{Massignan2008,Pricoupenko2010,JonaLasinio2010,Pricoupenko2011}.

\section{Determination of the model parameters}
In our model the scattering of two atoms is mediated by the exchange of the closed-channel or dimer field $\phi$. 
The two-body problem is thus solved by computing the renormalization of the inverse propagator of the dimer $\mathcal{G}_\phi^{-1}$. Evaluation of the standard ladder diagram yields
\begin{eqnarray}\label{eq:twobody}
\mathcal{G}_\phi^{-1}(E,\mathbf q) = P_\phi^\text{cl}(E,\mathbf q)-\frac{g^2/(32\pi)}{\sigma \left[1+\sigma\sqrt{-\frac{E}{2}+\frac{\mathbf q^2}{4}-i\epsilon}\right]^2}
\end{eqnarray}
with $P_\phi^\text{cl}(E,\mathbf q)=-E+\mathbf q^2/2-\nu(B)-i\epsilon$.  The two-atom scattering amplitude now follows from $f(k)=g^2\chi(k)^2\times$ $\mathcal{G}_\phi(2k^2,\mathbf 0)/(16\pi)$.  Its standard low-energy 
 expansion then determines the scattering length $a$ and the effective range $r_e$ via
\begin{equation}
\frac{1}{a}=\frac{1}{2\sigma}-\frac{16\pi}{g^2}\nu(B) ,\quad r_e=-2 r^*+3\sigma\left(1-\frac{4 \sigma}{3 a}\right).\label{effrange}
\end{equation}
This allows to express the bare parameters $g$, $\sigma$, and $\mu B_\text{res}$ which appear in \eqref{eq:action} in terms of 
fixed, experimental parameters. Close to a Feshbach resonance at magnetic field $B_0$, the scattering length can be written as 
$a(B)=-1/r^*\tilde\nu(B)$ where $\tilde\nu(B)=\mu(B-B_0)$ is the renormalized detuning in units of a wavenumber squared, 
while $r^{\star}>0$ is the intrinsic length scale which characterizes the strength of 
the Feshbach coupling~\cite{Chin2010,Bloch2008}. This fixes $g^2=32\pi/r^*$. Moreover, the resonance shift is given by $\mu(B_0-B_\text{res})=1/(r^{\star}\sigma)$, which is always positive in our model. This resonance shift has previously been calculated using microscopic
interaction potentials that have a van der Waals tail~\cite{Goral2004}. Comparison with this result yields the identification 
$\sigma=\bar a$, with the so-called mean scattering length $\bar a= 4\pi/\Gamma(1/4)^2 l_\text{vdw}\approx0.956\, l_\text{vdw}$ \cite{Gribakin1993}. 
All parameters of our model \eqref{eq:action} are thus fixed by two-body physics.

\section{Functional renormalization group solution of the three-body problem}
Based on the knowledge of the full two-body scattering amplitude, the three-body problem can be  
solved exactly, keeping only s-wave interactions. In particular, the three-boson scattering can be expressed in terms of an atom-dimer interaction $\sim \phi^*\psi^*\phi\psi$. The corresponding  one particle irreducible atom-dimer vertex 
$\lambda_3(Q_1,Q_2,Q_3)$ [$Q_i=(E_i,\mathbf q_i)$] develops a complicated energy and momentum dependence which determines the full Efimov spectrum for arbitrary values of the scattering length. The derivation becomes particularly simple using the functional renormalization group (fRG) \cite{Wetterich1993}. The central quantity of the fRG is an RG scale $k$ dependent effective action $\Gamma_k$ which interpolates between the microscopic action $S=\Gamma_{k=\Lambda}$ and the full quantum effective action $\Gamma=\Gamma_{k=0}$ by successively including quantum fluctuations on momentum scales $q\gtrsim k$. Here, we adopt an RG strategy adjusted to the few-body problem as discussed in \cite{Diehl2008,Moroz2009}, where the flowing action $\Gamma_k$ is of the form of $S$ in \eqref{eq:action} but with $P_\phi^\text{cl}$ replaced by $1/\mathcal{G}_\phi$ from \eqref{eq:twobody} and an additional three-body term
 \begin{figure}[t]
  \centering
  \includegraphics[width=\linewidth]{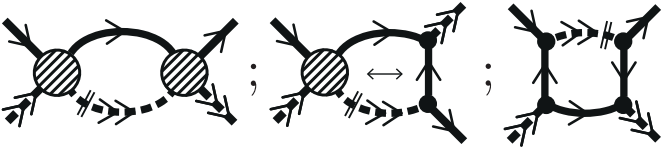} 
  \caption{\label{diagram} 
  Feynman diagrams contributing to the renormalization of the atom-dimer vertex $\lambda_3^{(k)}$ (large circle). The small black circle represents the atom-dimer coupling $\sim g$ and the solid (dashed) line denotes the atom (dimer) propagator.}
\end{figure}
\begin{eqnarray}\label{3Baction}
\Gamma_k^{\text{3B}}&=&-\int_{Q_1,Q_2,Q_3}\lambda_3^{(k)}(Q_1,Q_2,Q_3)\times\nonumber\\
&&\phi^*(Q_1)\psi^*(Q_2)\phi(Q_3)\psi(Q_1+Q_2-Q_3).
\end{eqnarray}
Since we do not consider a microscopic three-body force here, we have $\lambda_3^{(\Lambda)}=0$ at the UV scale $\Lambda$. The atom-dimer vertex $\lambda_3^{(k)}$ is then the only running coupling in $\Gamma_k$. It is important to note that the truncation of $\Gamma_k$ is complete for the solution of the three-body problem as no additional couplings can be generated in the RG flow \cite{Diehl2008,Moroz2009}. 
In our scheme the propagator of the bosons $\psi$ is not regularized and the dimer $\phi$ is supplemented with a sharp momentum regulator. In Fig. \ref{diagram} we show the Feynman diagrams contributing to the flow of $\lambda_3^{(k)}$. The number of independent energies and momenta is reduced by working in the center-of-mass frame and by noting that the loop frequency integration puts one internal atom on mass-shell \cite{Braaten2006}. 
After performing the s-wave projection $\lambda^{(k)}_3(q_1,q_2;E)=1/(2g)\int d\cos\theta \lambda_3^{(k)}(\mathbf q_1,\mathbf q_2;E)$, $\theta=\angle(\mathbf q_1,\mathbf q_2)$, one finds the RG equation
\begin{eqnarray} \label{flow1}
&&\partial_{k}\lambda_3^{(k)}(q_1,q_2;E)=-\frac{g^2 k^2\mathcal{G}_\phi(E-k^2,k)}{2\pi^2}\times\nonumber\\
&&\Big[\lambda_{3}^{(k)}(q_1,k;E)\lambda_3^{(k)}(k,q_2;E)+ \lambda_{3}^{(k)}(q_{1},k;E)G_E(k,q_{2})  \nonumber \\
&&+ G_E(q_{1},k)\lambda_{3}^{(k)}(k,q_{2};E)+ G_E(q_{1},k)G_E(k,q_{2}) \Big],
\end{eqnarray} where
\begin{equation}
G_E(p,q)\equiv \frac{1}{2}\int^{1}_{-1}d\cos\theta\frac{\chi(\left| \mathbf p+\frac{\mathbf q}{2}\right|)\chi(\left| \mathbf q+\frac{\mathbf p}{2}\right|)}{-E+\mathbf p^2+\mathbf q^2+(\mathbf p+\mathbf q)^2-i \epsilon}.
\end{equation}
Making use of the binomial form of Eq. \eqref{flow1} the flow can be integrated analytically and yields 
\begin{equation}\label{STM}
f_E(q_1,q_2)=g_E(q_1,q_2) - \int_0^\Lambda dl\, g_E(q_1,l)\, \zeta_E(l)\, f_E(l,q_2),
\end{equation}
which is a modified form of the well-known STM equation \cite{STM} with $f_E(q_1,q_2)= g_E(q_1,q_2)+ \tilde\lambda_E(q_1,q_2)$, $g_E(q_{1},q_{2})=16q_{1}q_{2}G_E(q_{1},q_{2})$, $\tilde\lambda_E(q_1,q_{2})=16q_{1}q_{2}\,\lambda_{3}(q_{1},q_{2};E)$,  and $\zeta_E(l)=-g^2\mathcal{G}_\phi(E-l^2,l)/(32\pi^2)$. \\
In a standard treatment of Efimov physics with contact interactions \cite{Braaten2006}, the STM equation \eqref{STM} has to be regularized and the resulting dependence on the UV cutoff scale $\Lambda$ reflects the presence of the three-body parameter. The exact position of the Efimov states is then adjusted by choosing an appropriate value of $\Lambda$. In our case, the situation is fundamentally different. Due to the presence of the form factor $\chi$ in $g_E$ and the finite range corrections in Eq. \eqref{eq:twobody}, the UV limit $\Lambda\to\infty$ can safely be taken and the usual three-body parameter completely disappears from the theory. As a result, the exact position of the states in the Efimov spectrum is predicted without any adjustable parameter.\\
The knowledge of the full vertex $\lambda_3$ gives all information about the scattering of three bosons, such as bound states, recombination rates, and lifetimes, by evaluating the corresponding tree-level diagrams \cite{Braaten2006}.  In the following we compute the trimer bound state spectrum by identifying the poles of $\lambda_3$ as a function of the energy $E$. As shown recently \cite{Stecher2009,Schmidt2010}, such 
poles also exist in higher order vertices $\lambda_N$
for $N\geq 4$ which are not considered in our work. 
Indeed, it is likely that there are $N$-body bound states
for arbitrary large $N$,\footnote{This has been shown numerically up to $N=13$, see \cite{Stecher2010}.} a conjecture consistent 
with a recent theorem by Seiringer \cite{Seiringer2012} which states that any
pairwise interaction potential with negative scattering 
length $a$ has an $N$-body bound state for some value 
of $N$, no matter how small $|a|$ may be. In the vicinity of a bound state pole the atom-dimer vertex can be parametrized as $\lambda_3(q_1,q_2;E)\approx \mathcal{B}(q_1,q_2)/[E+E^{(n)}+i \Gamma^{(n)}]$. When inserted into Eq.~\eqref{STM} an integral equation for $\mathcal{B}$ is obtained which is solved by discretization and amounts to evaluating the determinant $\det[\mathcal{C}-\mathbb{I}]=0$ with $\mathcal{C}(q_1,q_2)=g^2 g_E(q_1,q_2)\mathcal{G}_\phi(E-q_2^2,q_2)/(32\pi^2)$. 
$\mathcal{C}$ has a log-periodic structure where low-momentum modes are suppressed by any finite $1/a\neq0$ and energy $E<0$ below the atom-dimer threshold. High-momentum modes are suppressed due to the finite range potential of our model.

\section{Universal Efimov spectrum}
\begin{figure}[b!]
  \centering
  \includegraphics[width=\linewidth]{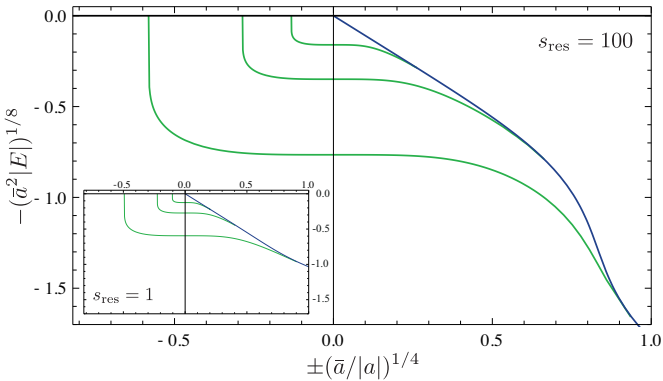} 
  \caption{\label{EfimovSpectrum} 
  (color online). The Efimov spectrum in dimensionless units for a open-channel dominated Feshbach resonance of strength $s_\text{res}=100$. The inset shows the spectrum for a resonance of intermediate strength $s_\text{res}=1$. The dimer binding energy is shown in blue.}
\end{figure}

\begin{figure}[t!]
  \centering
  \includegraphics[width=\linewidth]{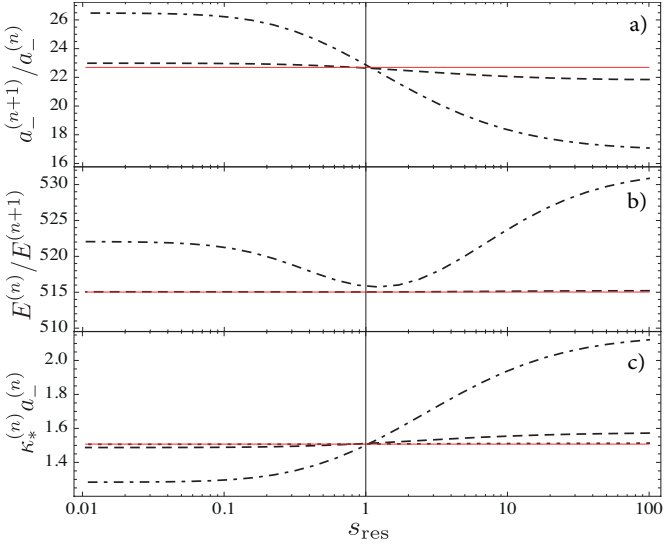} 
  \caption{\label{suppl} 
  (color online). Various dimensionless ratios as function of the Feshbach resonance strength $s_\text{res}$. Shown are the results for the lowest levels (black). The universal scaling result is shown in red. a) Ratio of scattering lengths $a_-^{(n+1)}/a_-^{(n)}$ where the consecutive trimer states meet the atom threshold at $E=0$. b) Ratio $E^{(n)}/E^{(n+1)}$ of the consecutive trimer energies at unitarity $1/a=0$.   c) Ratio $\kappa_*^{(n)}a_-^{(n)}$ as a measure of the distortion of the trimer levels from their universal shape in the $(a,E)$ plane. }
\end{figure}

\begin{figure}[t]
  \centering
  \includegraphics[width=\linewidth]{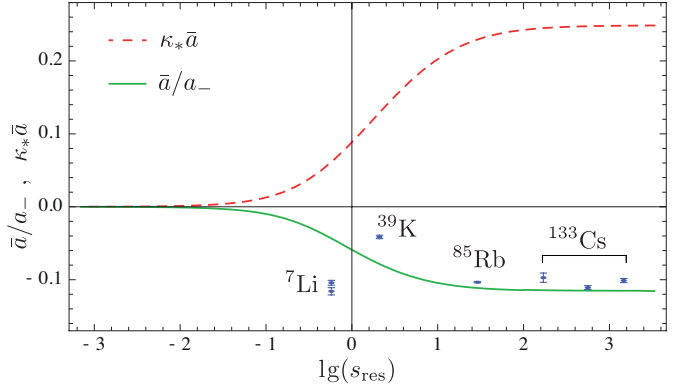} 
  \caption{\label{fig3} 
  (color online). Inverse threshold scattering length $a_-$ (solid line) and wavenumber $\kappa_*=\kappa^{(0)}_*$ (dashed line) in units of $\bar a$ as functions of the resonance strength $s_{\text{res}}$. The dots with error bars show the experimental results for $^7$Li \cite{Gross2010,Pollack2009}, $^{39}$K \cite{Zaccanti2009}, $^{85}$Rb \cite{Wild2011} and $^{133}$Cs \cite{Berninger2011}.}
\end{figure}

  In Fig.~\ref{EfimovSpectrum} we show the resulting Efimov spectrum including the atom-dimer threshold for an open-channel dominated Feshbach resonance 
  and one of intermediate strength in dimensionless units.  The position of the trimer states in the $(1/a,E)$ plane is completely fixed by our calculation. The overall appearance of the spectrum remains similar as the strength of the resonance is varied. In the limit $s_{\text{res}}\ll 1$, it gets pushed towards the 
  unitarity point $E=1/a=0$, while for open-channel dominated resonances it reaches a maximal extent in the $(1/a,E)$ plane.  
The detailed position of the lowest energy levels depends on both the value of the van der Waals length and the resonance strength $s_\text{res}$. Only in the experimentally hardly accessible limit, $n\gg 1$, the ratios of $a_-^{(n)}$,  $a_*^{(n)}$ (the scattering length for which the trimer meets the atom-dimer threshold),  and $E^{(n)}$ of consecutive levels approach their universal values.  It is instructive to quantify to which extent the lowest states deviate from this within our model. In Table \ref{table} and Fig. \ref{suppl} we show our results for various dimensionless ratios for Feshbach resonances of widely different strengths. Apparently, already for the third state the results are close
to the asymptotic behavior determined by the universal Efimov number  $s_0\approx1.00624$, regardless of the value of $s_\text{res}$.
\begin{table}[t!]
\scalebox{0.95}{
\begin{tabular}{ c c  r r r r}                    
 $s_\text{res}$&  	n	&	0	 &	 1 	& 	2  	& $n\gg1$	\\
   \hline \hline   
   \multirow{3}{*}{ \vspace{-5mm}100} 
  &$E^{(n)}/E^{(n+1)} $			& 530.871 	& 515.206		& 515.035		& 515.028 \\
  & $a_-^{(n+1)}/a_-^{(n)}$  		& 17.083	 	& 21.827	 	& 22.654		& 22.694 \\
  & $a_*^{(n+1)}/a_*^{(n)}$  		& 3.980		& 40.033		& 23.345 		& 22.694\\  
   &$\kappa_*^{(n)} a_-^{(n)}$ 		 & 2.121		& 1.573	 	& 1.512	   	& 1.5076 \\ 
  \hline  
   \multirow{3}{*}{ \vspace{-5mm}1} 
    &$E^{(n)}/E^{(n+1)} $			& 515.830 	& 515.039		& 515.035		& 515.028 \\
  & $a_-^{(n+1)}/a_-^{(n)}$  		& 22.869	 	& 22.650 		& 22.690		& 22.694 \\
  & $a_*^{(n+1)}/a_*^{(n)}$  		& 17.183		& 22.303		& 22.716	 	& 22.694\\  
   &$\kappa_*^{(n)} a_-^{(n)}$ 	 	& 1.500	 	& 1.511   	& 1.508		& 1.5076 \\ 		
   \hline
      \multirow{3}{*}{ \vspace{-5mm}0.1} 
  &$E^{(n)}/E^{(n+1)} $			& 521.273 	& 515.059		& 515.010		& 515.028 \\
  & $a_-^{(n+1)}/a_-^{(n)}$  		& 26.230 		& 22.964	 	& 22.71		& 22.694 \\
  & $a_*^{(n+1)}/a_*^{(n)}$  		& 26.965		& 21.286		& 22.48	 	& 22.694\\  
   &$\kappa_*^{(n)} a_-^{(n)}$ 		& 1.296  		& 1.489 		& 1.506	   	& 1.5076 \\ 
\end{tabular}
}
 \caption{\label{table}
  The ratio between consecutive trimer energies $E^{(n)}=\hbar^2 (\kappa_*^{(n)})^2/m$ and threshold scattering lengths ($a_{-,*}^{(n)}$) as well as the product $a_-^{(n)}\kappa_*^{(n)}$
for a open-channel ($s_\text{res}=100$), intermediate ($s_\text{res}=1$) and closed-channel dominated ($s_\text{res}=0.1$) Feshbach resonance for the three lowest-lying Efimov states $n=0,1,2$. The rightmost column shows the ratios in the universal scaling limit ($E=1/a=0$).}
\end{table}
By contrast, the experimentally most relevant lowest states exhibit large deviations. Remarkably, our prediction $a_{-}^{(1)}/a_{-}=17.08$ 
for open-channel dominated resonances is in reasonable agreement with recent measurements of the position of the second Efimov trimer in 
$^6\text{Li}$, which find a ratio near $19.7$~\cite{Wenz2009,Williams2009}, definitely smaller than the asymptotic value $22.69$.   
Surprisingly, for intermediate Feshbach resonances ($s_\text{res}\approx1$), the interplay between the scales $r^*$ and $\sigma$ leads to ratios close to their asymptotic ones even for the lowest states (see Fig. \ref{suppl}). Note that the values of $a_*^{(n)}$ for small $n$ are highly sensitive to the precise form of the two-body bound state spectrum which, on its own, is strongly non-universal. The ratios between the lowest $a_*^{(n)}$ are therefore in general not suitable for a measurement of universal ratios. Instead one has to access the states with $n\gg 1$ near threshold, where the dimer binding energy has the universal form $\epsilon_b=\hbar^2/ma^2$.\\
To investigate generic features of the trimer spectrum which are independent of the precise form of the dimer energy, we study the dependence of $a_-^{(n)}$ and $\kappa_*^{(n)}$ on the strength of the Feshbach resonance. In Fig.~\ref{fig3} the behavior for the lowest, experimentally accessible, state is shown. For open-channel dominated resonances $a_-/\bar a$ and $\bar a\kappa_*$ become independent of  $s_\text{res}$ and thus of $r^*$ and we find $a_-\approx-8.27\, l_\text{vdw}$ and $\kappa_*l_\text{vdw}=0.26$. In the limit of closed-channel dominated resonances, the van der Waals length becomes irrelevant and the scale for the full Efimov spectrum is set by $r^*$ only. Specifically, we find $a_-^{(n)}=\xi^{(n)} r^*$ and $\kappa_*^{(n)}r^*=\eta^{(n)}$ with numbers $\xi^{(n)}$ and $\eta^{(n)}$ which approach universal values as $n\to\infty$. In fact, we accurately reproduce the results for the universal scaling limit ($n\gg1$) of closed-channel dominated Feshbach resonances,  $a_-(n\gg1)=-12.90\, r^*$ and $\kappa_*(n\gg1)r^*=0.117$,  which were previously derived within a zero range model where $\sigma=0$ \cite{Petrov2004,Gogolin2008}.  The low-lying Efimov states however deviate from this limiting scaling behavior and our model predicts for example for the lowest Efimov state the ratios $a_-=-10.3\, r^*$ and $\kappa_*r^*=0.125$. The precise numbers which quantify the deviations from the universal scaling predictions are specific for our model \eqref{eq:action} with an exponential form factor $\chi(r)$. Note that it is also possible to study these deviations using a systematic expansion in the small parameter $l_{vdw}/|a|$ within effective field theory as done by Ji and coworkers \cite{Ji2010}. Such an approach, however, requires not only an adjustable three-body parameter, but in addition further counterterms which are necessary to renormalize the theory with an interaction of finite range at finite scattering lengths $a$.

\section{Comparison to experiments}
Comparing to the experimental data, the open-channel dominated resonances in $^{85}\text{Rb}$ \cite{Wild2011} and in $^{133}\text{Cs}$ \cite{Berninger2011} fit well into our prediction, see Fig.~\ref{fig3}. The $12\,\%$ deviation between the
value $a_-\approx-9.45\, l_\text{vdw}$ inferred from averaging the results from different experiments and our ratio $-8.27$ shows that there is
no true universality of the 3BP in a strict sense: short range physics which enters into the details of the form factor $\chi(r)$ leads to slightly different numbers.  From both the empirical data and a study of how much our numbers change 
for various choices of the form factor $\chi(r)$, they are generically at the ten percent level.  As mentioned above,  the van der Waas length does not set the scale for the Efimov spectrum in general. 
In particular, in the regime of Feshbach resonances of intermediate strength 
$s_\text{res}\approx 1$ both scales 
$r^*$ and $\sigma$ become relevant. Our approach equally applies to this regime, which is realized, e.g., in the case of $^{39}\text{K}$, where 
$s_\text{res}\simeq 2.1$~\cite{Zaccanti2009}. As shown in Fig.~\ref{fig3}, the observation~\cite{Zaccanti2009} of a considerable deviation from the 
result  $a_-\!\approx\! -9.45\, l_{\text{vdw}}$ in this case is in qualitative agreement with our model\footnote{note added: new experimental data suggests that the value of $l_{\text{vdw}}/a_-$ reported in \cite{Zaccanti2009} has to be corrected and is shifted to a larger value \cite{privcom}.}.  
By contrast, the case of  $^7\text{Li}$, which seems to follow nicely the result $a_-\!\!\approx\!\! -9.45\, l_{\text{vdw}}$~\cite{Pollack2009,Gross2010} 
for open-channel dominated resonances despite the even smaller value $s_\text{res}\simeq 0.58$~\cite{Gross2010} of the resonance strength is 
\textit{not} consistent with our prediction.  A possible origin of this discrepancy may be three-body forces of the Axilrod-Teller type~\cite{Axilrod1943}, which lead to substantial changes in the position of the first Efimov trimer~\cite{DIncao2009}. Their magnitude, in fact, depends 
quite sensitively on the choice of how these forces are modeled at short distances. Whether it is indeed three-body forces or other
effects not captured by our model that will account for the discrepancy between the observed $a_-$ in  $^7\text{Li}$ and our 
prediction is unknown at present. Note, however, that irrespective of this problem, an explanation of the observed result for $a_-$ in 
 $^7\text{Li}$ within a single-channel description~\cite{Greene2012} is likely to be inadequate due to the rather small value of $s_\text{res}\simeq 0.58$.
 
\section{Conclusion}
We have presented a simple, exactly solvable model, containing only $r^*$ and $l_\text{vdw}$ as experimentally accessible parameters, 
in which the full Efimov spectrum is fixed in quantitative terms without an adjustable 3BP. Our results provide an explanation for the observed 
proportionality between the scattering length $a_{-}$ where the first Efimov trimer appears and the van der Waals length $l_\text{vdw}$, which is often interpreted as a 'universality' of the 3BP. This relation applies for open-channel dominated resonances and in situations in which three-body forces are negligible. A continuous crossover is found into the regime of closed-channel dominated resonances, where the scale for the 3BP is set by $r^{\star}$, recovering previous exact solutions~\cite{Petrov2004,Gogolin2008}. Our results provide a clue for why the ratio  $a_{-}/l_\text{vdw}$ in $^{39}\text{K}$ 
is quite different from those in the open-channel dominated resonances. It remains an open question, however, why the $^7\text{Li}$ resonance, which 
is far from being open-channel dominated, has an $a_{-}$ that fits perfectly the open-channel dominated limit.
Finally, we have shown that for the lowest Efimov states within a given 
trimer spectrum, there are appreciable deviations from the asymptotic scaling relations, consistent with experiments. 
Clearly, a more systematic investigation of these non-universal ratios and of resonances with intermediate strength 
$s_\text{res}=\mathcal{O}(1)$ is necessary to clarify to which extent the generic features of the Efimov effect in ultracold atoms 
are captured by our simple model, in which the complete trimer spectrum is obtained without any adjustable parameter from two-body physics only.

\begin{acknowledgement}
  We thank Francesca Ferlaino, Rudi Grimm, Selim Jochim, Robert Seiringer, Felix Werner and Matteo Zaccanti for useful discussions and acknowledge support by the DFG through FOR 801.
\end{acknowledgement}
%
%

\end{document}